\documentclass[aps,prd,preprint,superscriptaddress,nofootinbib,showpacs]{revtex4}
\usepackage{graphics}
\usepackage{epsfig}
\usepackage{latexsym}
\usepackage{colordvi}
\usepackage{amsmath}
\usepackage{amssymb}

\newcommand{\be}{\begin{equation}}
\newcommand{\ee}{\end{equation}}
\newcommand{\bea}{\begin{eqnarray}}
\newcommand{\eea}{\end{eqnarray}}

\newcommand{\p}{\partial}

\newcommand{\nn}{\nonumber}

\def\half{\frac{1}{2}}

\def\IB{\relax\hbox{$\inbar\kern-.3em{\rm B}$}}
\def\IC{\relax\hbox{$\inbar\kern-.3em{\rm C}$}}
\def\ID{\relax\hbox{$\inbar\kern-.3em{\rm D}$}}
\def\IE{\relax\hbox{$\inbar\kern-.3em{\rm E}$}}
\def\IF{\relax\hbox{$\inbar\kern-.3em{\rm F}$}}
\def\IG{\relax\hbox{$\inbar\kern-.3em{\rm G}$}}
\def\IGa{\relax\hbox{${\rm I}\kern-.18em\Gamma$}}
\def\IH{\relax{\rm I\kern-.18em H}}
\def\IK{\relax{\rm I\kern-.18em K}}
\def\IL{\relax{\rm I\kern-.18em L}}
\def\IP{\relax{\rm I\kern-.18em P}}
\def\IR{\relax{\rm I\kern-.18em R}}
\def\IZ{\relax{\rm Z\kern-.5em Z}}

\begin{document}
\preprint{\parbox[b]{1in}{ \hbox{\tt PNUTP-09/A02}}}

\title{Spin 3/2 Baryons and Form Factors in AdS/QCD}

\author{Hyo Chul Ahn}
\affiliation{Department of
Physics, Pusan National University,
             Busan 609-735, Korea}

\author{Deog Ki Hong}
\affiliation{Department of
Physics,   Pusan National University,
             Busan 609-735, Korea}

\author{Cheonsoo Park}
\affiliation{Department of
Physics, Pusan National University,
             Busan 609-735, Korea}

\author{Sanjay Siwach}
\affiliation{Department of
Physics, Pusan National University,.
             Busan 609-735, Korea}
\affiliation{Department of
Physics,  Banaras Hindu University,
Varanasi-221005, India\footnote{permanent address}}

\date{\today}

\begin{abstract}

We study the 5D Rarita-Schwinger fields to describe
spin 3/2 baryons in AdS/QCD.
We calculate the spectrum of spin 3/2 baryons ($\Delta$ resonances) and their form factors, together
with meson-baryon couplings from AdS/QCD.
The transition form-factors between $\Delta$ and nucleon
are evaluated. Both pion and rho meson couplings have the same origin in
the bulk and hence unified. The numerical values for the
meson-baryon transition couplings are consistent with the values obtained
from other methods. We also predict the numerical values of some new
couplings associated with $\Delta$ resonances.

\end{abstract}
\pacs{14.20.Gk,13.40.Gp,11.25.Tq,11.10.Kk}

\maketitle


\section{Introduction}
Solving Quantum Chromodynamics (QCD) is a long outstanding problem in physics.
Recently
gauge-gravity duality or Anti-de Sitter/Conformal Field Theory (AdS/CFT) correspondence~\cite{Maldacena:1997re, Witten:1998qj} has been used to study
the strong coupling regime of gauge theories from string theory. The
qualitative and quantitative results obtained in this manner
indicate that we may hope to get QCD from string theory.

AdS/CFT correspondence may be adapted to describe the low energy
dynamics of QCD. In the simplest model known as AdS/QCD, the AdS
spacetime is cutoff at a finite distance from the ultraviolet  boundary in
AdS/CFT~\cite{Erlich:2005qh,deTeramond:2005su}. For related works,
see also~\cite{Son:2003et}. This amounts to placing an infrared
brane at some position in the bulk of AdS spacetime. The hadronic
spectra obtained from AdS/QCD is 10-15 $ \% $ off the experimental
values. This hints that AdS/QCD may be a tip of some more accurate
description of QCD from string theory. It has been shown recently
that the results for the QCD current-current two point point
functions from AdS/QCD are related to Migdal's approach to
hadronization~\cite{Erlich:2006hq}. In mathematical terms Migdal's
approach amounts to Pade approximation. The other models of QCD from
string theory~\cite{Kruczenski:2003uq,Sakai:2004cn} use the
microscopic description in terms of intersecting D-branes in string
theory and perhaps AdS/QCD is some limit of these models. In the
D-brane setup the baryons are realized as instanton solitons of the
low energy effective
theory~\cite{Nawa:2006gv,Hong:2007kx,Hata:2007mb}.

In this paper we shall be interested in the spectrum of spin 3/2
baryons ($\Delta$ resonances) and  the couplings between mesons and
baryons from AdS/QCD. The holographic description of spin 3/2
operators in the boundary theory is given by the Rarita-Schwinger
fields in the bulk~\cite{Volovich:1998tj,Hong:2006ta}. The spectrum of
spin 3/2 baryons is obtained by the eigenvalue equations for the normalizable modes of
bulk Rarita-Schwinger fields, subject to appropriate boundary conditions
at both the infrared (IR) and ultraviolet (UV) branes, where the  IR brane is  introduced by hand
to implement the confinement of quarks, setting the scale for AdS/QCD, while the UV brane will be located infinitesimally close to origin.
In AdS/QCD the chiral symmetry breaking is encoded in the vacuum expectation value of
bulk scalar, dual of the order parameter of chiral symmetry at the boundary.

The effect of Yukawa couplings on the
spectrum is very small. We also find the numerical results for the
coupling of pions and rho mesons with nucleons and spin 3/2 baryons.
The pions and rho meson couplings arise from the same terms in the
bulk and this allows us to predict some couplings not known so far.
The transition form factors between nucleons and $\Delta$ are evaluated numerically for various values of momentum $Q^2$.

The plan of the paper is as follows. In the next section we briefly
overview the chiral dynamics and the spectrum of mesons from AdS/QCD
\cite{Erlich:2005qh}. We also discuss the spectrum of nucleons
\cite{Hong:2006ta}. In next section we consider the spectra of spin
3/2 baryons. The effect of bulk Yukawa couplings is also considered.
In the following two sections we calculate the meson-baryon transition
couplings and the nucleon-$\Delta$ transition form factors
from AdS/QCD. The last section is devoted to discussion
and conclusions.

\section{Mesons and Nucleons in AdS/QCD}

We shall be interested in the low-energy chiral dynamics of hadrons
from AdS/QCD, proposed as the holographic dual of QCD~\cite{Erlich:2005qh}. The global flavor currents in
the boundary theory will correspond to the bulk gauge fields, according to AdS/CFT correspondence. The bulk scalar
$X$, which transforms as a bifundamental under chiral symmetry group
$SU(2)_L \times SU(2)_R$, acts as an order parameter for the chiral
symmetry breaking. In the boundary theory the expectation value of
its dual operator, $\bar q_L q_R$ can be identified with the quark
condensate. The minimal action in the bulk suitable for chiral
dynamics can be written as
\bea S_{\rm AdS} = \int dz d^4 x\,
\sqrt{G}\,{\rm Tr} \Big[\left|DX\right|^2 - M^2_5 \left|X\right|^2 -
\frac{1}{4g^2_5}(F^2_L + F^2_R) \Big] \eea where $G$ is the
determinant of 5-dimensional AdS metric, \bea ds^2 =
\frac{1}{z^2}\Big(-dz^2 + \eta_{\mu\nu}dx^\mu dx^\nu\Big), \eea and
$\eta_{\mu\nu}$ being the flat 4-dimensional metric with the
signatue $(+,-,-,-)$. The fifth coordinate $z$ is cut-off at the
infra-red scale $z_m = 1/\Lambda_{QCD}$, corresponding to the
confinement scale of QCD,  and one has to regulate the bulk action
at the ultraviolet (UV) boundary $z = \epsilon \rightarrow 0$. The
covariant derivative is defined as $D_M X = \p_M X - i A_{LM}\,X + i
X\,A_{RM}$. The five dimensional gauge coupling $g_5$ is related to
the rank of the gauge group (number of colours) in the boundary
theory \cite{Erlich:2005qh} and we take $g_5 = 2 \pi$ (for $N_c = 3$
and for two flavors). The five dimesninal mass of the scalar, $X$,
is fixed by the scaling dimension, $\Delta_{0}$,  of the operator $\bar q_L q_R$ in
the boundary theory as $ M^2_5 =
\Delta_{0}(\Delta_{0}-4)$ in the unit of inverse AdS radius~\cite{Gubser:1998bc,Witten:1998qj}.

The classical solution for the bulk scalar $X$ can be written as,
\bea \left<X(z)\right> = \half M z + \half \sigma z^3. \eea By the
AdS/CFT duality, the coefficient of the non-normalizable mode is
identified as the source for the boundary operator $\bar q_L q_R$
and the coefficient of the normalizable mode as its vacuum expectation value,
$\sigma=\left<{\bar q_{L}}q_{R}\right>$.  In the chiral limit,
$M\rightarrow 0$, therefore we can write $X = v(z)\, e^{i\,P(x,z)}$
with $v(z) = \half \sigma z^3$ and $P(x,z)$ is proportional to the
pion fields upon the Kaluza-Klein reduction to four dimensions and
$\sigma $ becomes the order parameter for the chiral symmetry
breaking. Then one can identify the pions as the Nambu-Goldstone
bosons of the broken symmetry.

In the leading approximation, one can work with linearized equations
for the gauge fields. In the unitary gauge a linear combination
$P(x,z)$ and the fifth component of axial gauge fields $A_z$,
$z\partial_z(A_z/z)-2v^2g_5^2P/z^2$, and the fifth component of the
vector gauge field become infinitely massive and  decouple from the
theory, while the orthogonal combination of $P$ and the axial gauge
fields remain as a physical degree of freedom~\cite{Hong:2006ta} if
\begin{eqnarray}
z\partial_z\left(\frac{A_z}{z}\right)-2\frac{v^2g_5^2}{z^2}P=0\,,
\label{ortho}
\end{eqnarray}
which then relates  pions with the fifth component of axial gauge field, $A_z$.
The
four dimensional components of the vector and axial vector bulk gauge
fields can be identified with the infinite tower of vector and axial-vector mesons
respectively, upon Kaluza-Klein reduction.

Next we turn to the description of baryons in AdS/QCD, as studied in~\cite{Hong:2006ta}. To describe
spin 1/2 baryons in accordance with AdS/CFT correspondence,  we introduce a bulk spinor whose normalizable modes will correspond
to  color-singlet  spin-1/2 states in the boundary theory (QCD), while its non-normalizable modes will couple to those states as sources.  A 5D bulk spinor has same degrees of freedom as 4D Dirac spinor, which contains both left-handed components and right-handed components, transforming differently under (flavor) chiral symmetry of QCD. Since the flavor symmetry of boundary theory corresponds to gauge symmetry in the bulk, we need a pair of spinors (say $N_1$ for the left-handed components and $N_2$ for the right-handed components) in the bulk to describe a single Dirac spinor of boundary theory.

Furthermore, to calculate the correlation functions of baryons, we need to introduce sources in the boundary that couple to baryons as
\begin{equation}
\int\,{\rm d}^{4}x\left(\bar {\cal O}_{L}B_{R}+\bar B_{L}{\cal O}_{R}+{\rm h.c.}\right)\,.
\end{equation}
Since the source that couples to a chiral component of baryon has an opposite chirality,
the non-normalizable mode of $N_{1} (N_{2})$ becomes the source of the normalizable mode
of $N_{2} (N_{1})$\,.

The minimal bulk action for  the
spinors can be written as,
\begin{equation}
S_{N} = \int dz d^4 x \sqrt{G} \Big[i\bar N_1 e_{A}^{M}\Gamma^A D_M N_1
- m_5 \bar N_1 N_1 + (1 \leftrightarrow 2\,\, \&\,\,
m_5\leftrightarrow -m_5) \Big],
\end{equation}
where $e_{M}^{A}$ are the 5D vielbein, $e_{M}^{A}e_{N}^{B}\eta_{AB}=g_{MN}$, and $\Gamma^{A}$ ($A=0,1,2,3,5$)  are the $4\times4$ gamma matrices in 5D and the gauge and
Lorentz covariant derivative $D_{M}$ is given by
\begin{equation}
D_M=\partial_M-\frac{i}{4}\omega_{M}^{AB}\Sigma_{AB}-i(A^a_L)_Mt^a,
\end{equation}
where $\omega_{M}^{AB}$ is the spin connection and $\Sigma_{AB}=\frac{1}{2i}[\Gamma_A\,,\Gamma_B]$.
The bulk spinor
mass $m_5$ is fixed up to a sign by the conformal dimension, $\Delta_{1}$ of the
boundary spinor as $ m^2_5 =(\Delta_{1}-2)^{2}$ and the sign of spinor masses are chosen such that it preserves 4D parity~\cite{Hong:2006ta}.

When the interactions are turned off, the spinors $N_1$ and $N_2$
obey free Dirac equations in the bulk. To analyze the Dirac equations, it is convenient to use the chirality basis,
\begin{equation}
\Gamma^{5} =-i\gamma^{5}=\begin{pmatrix}
-i & 0\\
0 & i
\end{pmatrix}\,,
\quad \Gamma^{0}=\begin{pmatrix} 0& -1 \\ -1 & 0 \end{pmatrix}\,,
\quad \Gamma^{i}=\begin{pmatrix} 0& \sigma^{i} \\ -\sigma^{i} & 0 \end{pmatrix}\,,
\quad (i=1,2,3)\,.
\end{equation}
We first decompose the 5D spinor as
\begin{equation}
N_{i}(x,z)=N_{iL}(x,z)+N_{iR}(x,z), \quad (i=1,2)\,,
\end{equation}
where $\gamma^5N_{iL} =
N_{iL}$ and $\gamma^5N_{iR} =- N_{iR}$.

Let us consider the (free) Dirac equation for the spinor
$N_1$ first,
\begin{equation}
ie^{M}_{A}\Gamma^{A}\left(\partial_{M}-\frac{i}{4}\omega_{M}^{BC}\Sigma_{BC}\right)\,N_{1}-m_{5}N_{1}=0\,.
\end{equation}
Using the local Lorentz symmetry, we take $e_{M}^{A}=\frac{1}{z}\eta^{A}_{M}$. Then the only non-vanishing components of the spin connections are
\begin{equation}
\omega_{\mu}^{5A}=-\omega_{\mu}^{A5}=\frac{1}{z}\delta_{\mu}^{A}\quad (\mu=0,1,2,3)\,.
\end{equation}
The Dirac equations become
\begin{equation}
\left(z\gamma^{5}\partial_{z}+zi\!\not\!\partial\,-2-m_{5}\right)N_{1}=0\,.
\end{equation}
It is convenient to Fourier-transform the bulk spinor as
\begin{equation}
N_{iL,R}(x,z)=\int_{p}f_{iL,R}(p,z)\psi_{L,R}(p)e^{-ip\cdot x}\,,
\end{equation}
where the 4D spinors satisfy
\begin{equation}
\not\!p\,\psi(p)=|p|\psi(p)\,,
\end{equation}
where $|p|=\sqrt{p^{2}} $ for a time-like four-momentum $p$.
The corresponding Dirac equation can be written as,
\bea
 \left(\p_z - \frac{2 + m_5}{z}\right) f_{1L}&=& - |p|~f_{1R}, \nn\\
 \left(\p_z - \frac{2 - m_5}{z}\right) f_{1R}&=& |p|~f_{1L}\,,
\eea
Near the UV boundary, $z=\epsilon$,  we find
\begin{equation}
f_{1L}\simeq c_{1}(1+2m_{5})z^{2+m_{5}}+c_{2}|p|
z^{3-m_{5}}\,,\quad
f_{1R}\simeq c_{1}|p|z^{3+m_{5}}+c_{2}(2m_{5}-1)z^{2-m_{5}}\,
\end{equation}
where $c_{1},c_{2}$ are constants to be fixed by boundary conditions.

We first note that the spin-1/2 baryons should have massless modes in the spectrum
to satisfy  the 't Hooft anomaly matching condition
when the chiral symmetry is restored or baryons do not couple to the chiral symmetry breaking order parameter. In AdS/QCD the bulk spinors should have normalizable zero modes ($|p|=0$)
when the interactions are turned off.
We choose $m_5 > 0$ for $N_1$ to be consistent with the chirality of
the operator in the boundary theory, which fixes the boundary
conditions at UV and IR
\begin{equation}
f_{1L }(p,\epsilon) = 0, ~~~~~ f_{1R} (p,z_m) = 0.
\end{equation}
The remaining boundary conditions for $f_{1L} (x,z_m)$ and $f_{1R}
(x,\epsilon)$ follow from the equations of motions, coming from the boundary term. The normalizable
solutions for non-zero modes ($|p|\ne0$) are given by
\begin{equation}
f_{1L,R}(p,z) \sim  z^{5/2} J_{m_5\mp\half} (|p|z).
\end{equation}
The free spectrum with the boundary condition $N_{1R} (x,z_m) = 0$ is
given by the zeros of the Bessel function $J_{m_5 + \half} (|p| z_m)$.
Since the AdS/CFT gives the relation, $\Delta = |m_5| +
2$ for Dirac fields, we have  $m_5 = 5/2$ for the canonical dimension $\Delta =
9/2$ if we neglect the anomalous dimension of spin-1/2 baryons.

Similarly one can discuss the bulk spinor $N_2$ whose normalizable modes correspond to the right-handed components of baryons at boundary.  The sign of mass
term for $N_{2}$ is opposite to that of $N_{1}$ and the boundary
conditions are replaced by $N_{2R}(x,\epsilon)=0$ and $N_{2L} (x,z_m) = 0$.
Then one has the same
spectrum as for the left-handed baryons, as it should be for QCD.

In QCD we know that the bulk of baryon mass comes from chiral-symmetry breaking. Since the condensate $\left<\bar q\,q\right>$ is the order parameter of chiral symmetry breaking, the baryons should get mass through coupling to the condensate. In AdS/QCD~\cite{Hong:2006ta} it was shown that this can be easily achieved by introducing
Yukawa couplings between bulk spinors and bulk scalars,
\begin{equation}
{\cal  L}_{Yukawa} = - g \left(\bar N_2 X N_1 +\bar N_{1}X^{\dagger}N_{2}\right).
\end{equation}
The zero modes now acquire the mass via Yukawa couplings. Also the
degeneracy of excited states between parity-even and parity-odd baryons is lifted by this term. One can
estimate $g$ and $z_m$ by fitting the mass spectrum for the lowest-lying state (with the experimental value $938\,{\rm  MeV}$), which gives $g
= 14.4$ and $z_m^{-1} = 205\,{\rm  MeV}$. Then one can predict the mass
spectrum for excited states and numerical values agree well with the
experimental results. Furthermore, one can show that the pion coupling with excited baryons becomes smaller for more excited states such that chiral symmetry is effectively restored for highly excited states~\cite{Hong:2006ta}.

\section{Spin 3/2 Baryons ($\Delta$ resonances)}

Next to lowest-lying baryons are $\Delta$ resonances, which have spin 3/2 and also isospin 3/2.
The ground-state $\Delta$ baryon has even parity and its mass is measured to be $1210\,{\rm MeV}$ with decay width about $100\,{\rm MeV}$. Being the closest resonance to nucleons,  $\Delta$ baryons are very important in studying the properties of nucleons such as nucleon potential, as the dominant decay channel of $\Delta$ resonance is $\Delta\to N\pi$. Recently the transition form-factors of $\Delta$-to-nucleons are accurately measured~\cite{Stuart:1996zs} and also studied intensively in lattice~\cite{Alexandrou:2007zz,Alexandrou:2007dt} and in chiral perturbation theory~\cite{Procura:2008ze}. In this paper we investigate the properties of $\Delta$ resonances in AdS/QCD, including the $\Delta$-nucleon transition form-factors. It will be interesting to compare our results with the experimental data and also with the lattice calculations.

A simple holographic description of spin 3/2 baryons in the boundary
theory is given by the Rarita-Schwinger fields $\Psi_{M}$ in the bulk
\cite{Volovich:1998tj}. The action for the Rarita-Schwinger field in
$\text{AdS}_5$ is given by
\begin{equation}\label{RS}
\int
d^5x\sqrt{G}\left(i\bar{\Psi}_A\Gamma^{ABC}D_B\Psi_{C}-m_1\bar{\Psi}_A\Psi^A-m_2\bar{\Psi}_A\Gamma^{AB}\Psi_B\right)\,,
\end{equation}
where $\Psi_{A}=e_{A}^{M}\Psi_{M}$ and we used notations
$\Gamma^{ABC}=\frac{1}{3!}\Sigma_{\rm{perm}}(-1)^p\Gamma^A\Gamma^B\Gamma^C=\frac{1}{2}(\Gamma^B\Gamma^C\Gamma^A
-\Gamma^A\Gamma^C\Gamma^B)$ and
$\Gamma^{AB}=\frac{1}{2}[\Gamma^A\,,\Gamma^B]$. The Rarita-Schwinger
equations in $AdS_5$ are then written in the following form,
\bea
i\Gamma^A\Big(D_A \Psi_B - D_B \Psi_A\Big) - m_{-} \Psi_B + \frac
{m_{+}}{3}\Gamma_B\Gamma^A\Psi_A = 0\,,
\eea
where $m_\pm = m_1 \pm
m_2$. The values of $m_1$ and $m_2$ correspond to those of spinor
harmonics on $S^5$ of $\text{AdS}_5 \times S^5$ \cite{Kim:1985ez}.

Being a reducible  (axial) vector-spinor,
the Rarita-Schwinger fields contain not only spin-3/2 components but also spin-1/2 components as well.
In 4D, the extra spin-1/2 components can be projected out by a Lorentz-covariant constraint,
\begin{equation}
\gamma^{\mu}\Psi_{\mu}=0\,.
\end{equation}
As in 4D, the following Lorentz-covariant constraint will project out one of the spin-1/2 components from the 5D Rarita-Schwinger fields
\begin{equation}
e_{A}^{M}\Gamma^{A}\Psi_{M}=0\,,
\end{equation}
which then gives $\partial^{M}\Psi_{M}=0$ for a free particle if combined with equations of motion.

The 5D Rarita-Schwinger fields have one more extra spin-1/2 components, $\Psi_{z}$, if reduced to 4D.
In the case of bulk vector fields, the 5th component  constitutes the longitudinal component of massive spin-1 vector mesons of boundary theory or can be gauged away in the unitary gauge.
However, in the case of bulk Rarita-Schwinger fields there is no gauge degree of freedom but we may
choose  $\Psi_z = 0$ to further reduce the extra spin-1/2 degrees of freedom, because there is no boundary extra spinor that it can be mapped into, and convert the Rarita-Schwinger equations to a set of Dirac
equations for the remaing components,
\begin{equation}
\Big(iz\Gamma^A\p_A +2i \Gamma^{5} - m_{-} \Big) \Psi_\mu = 0,\quad (\mu=0,1,2,3)\,.
\end{equation}

To describe the spin 3/2 baryons ($\Delta$ resonances) in AdS/QCD
one has to introduce a pair of Rarita-Schwinger fields in the
bulk, $\Psi_1^A$ (for the left-handed spin-3/2) and $\Psi_2^A$ (for the right-handed spin-3/2), which obey the above
Rarita-Schwinger equations. Similarly to the case of spin-1/2  in the previous section,
using the 4D Fourier decomposition of the 5D spinors in the basis of chirality,
\begin{equation}
\Psi^A_{iL,R}(x,z)=\int_{p}F_{iL,R}(p,z)\psi^A_{L,R}(p)\,e^{-ip\cdot x}\,,
\end{equation}
we can obtain
the solutions for spin-3/2 baryons. However, unlike with the spin-1/2 spinor, we should impose boundary conditions such that  there are no zero modes
for spin 3/2 fields, since the anomalies are already saturated by spin-1/2 baryons. The normalizable solutions for non-zero modes
are given by,
\begin{equation}
F_{1L,R}(p,z) \sim z^{5/2} J_{m_-\mp\half} (|p|z).
\end{equation}
The boundary condition suitable for the description of left handed
spin 3/2 baryons is,
\begin{equation}
\Psi^M_{1R} (x,z_m) = 0.
\end{equation}

The free spectrum with the above boundary condition is given by the
zeross of the Bessel function $J_{m_- - \half} (|p| z_m)$.  As the AdS/CFT gives the relation
$|m_-| =\Delta_{3/2}- 2$ for Rarita-Schwinger fields of scaling dimension $\Delta_{3/2}$, we get $|m_-| = 5/2$ for $\Delta_{3/2} = 9/2$.
For the right-handed components of spin 3/2 baryons we have the similar
boundary condition but with $L
\leftrightarrow R$ as the sign of mass term for $\Psi_{2}^{M}$ has to be changed, $m_- \leftrightarrow -m_-$, giving the same mass spectrum.

To see the effects of chiral symmetry breaking, we consider the Yukawa couplings,
\begin{equation}{\cal  L}_{Yukawa} = - g_{3/2}\, \bar \Psi^M_2 X^3 \Psi_{1M} + \text{h.c.}\,.
\end{equation}
We find that the effect of Yukawa coupling is very small for spin
3/2 baryons for a sizable value of $g_{3/2}$ as the parameter $\sigma$,
which is very small, appears with third power in Yukawa couplings.
Using a cut off $z_m^{-1} = 205 {\rm MeV}$ (same as in the nucleon case) and taking $g_{3/2}=215$, we obtain  the $\Delta$ resonance masses as
\begin{equation}
0.90^{(1232)},~~~ 1.39^{(1700)},~~~ 1.67^{(1920)} ~~~({\rm GeV}),
\end{equation}
which differ by 13-27 $\%$ from the experimental results, denoted as superscripts.

\section{Meson-Baryon transition couplings}

The meson-baryon transition couplings can be evaluated in the framework of AdS/QCD.
The most interesting meson-baryon couplings are of pions and rho
mesons with the nucleons. They arise from the same term in the bulk
and hence unified.  The couplings of pions and rho meson with
baryons can be read off from the chiral Lagrangian. The lowest order
terms in the 4-D chiral Lagrangian give the couplings of pions with
nucleons as,
\bea
{\cal L}_{\pi N N} = - \frac{g_A}{f_\pi}  \bar\psi
\gamma^\mu \gamma^5 \p_\mu \pi \psi, \eea where $\pi(x) = \pi^a
\tau_a$ is the pion field and $g_A$ and $f_\pi=93\, \text{MeV}$ being the
nucleon axial couplings and the pion decay constant, respectively.

Using the equations of motion for $\psi$, the leading term can be
written as, \bea
 2 i g_{\pi N N} \bar\psi \gamma^5 \pi \psi,
\eea where $g_{\pi N N} = g_A m_N/f_\pi$ is the Goldberger-Treiman
relation and $m_N$ is the nucleon mass. The experimental value of
pion-nucleon coupling is measured to be  $g_{\pi N N} \simeq 13.5$ from the
pion-nucleon scattering. The couplings of rho mesons with nucleons
are given as
\bea
{\cal L}_{\rho N N} =  g_{\rho NN} \bar\psi
\gamma^\mu V_\mu \psi + \frac{f_\rho}{4m_N} \bar\psi \sigma^{\mu\nu}
(\p_\mu V_\nu - \p_\nu V_\mu) \psi, \eea where the second term is
the magnetic type of coupling, known as nuclear tensor coupling.

As the $\Delta$ resonance is parity-even and described by an axial-vector spinor, the couplings of pions and rho mesons with nucleons and $\Delta$  are given by
\begin{equation}
{\cal L}_{\pi N \Delta} = - i \frac{g_{\pi N \Delta}}{f_\pi} \bar
\psi^\mu\p_\mu \pi \psi+i g_{\rho N \Delta} \bar\psi^\mu \gamma^5
V_{\mu} \psi.
\end{equation}
Similarly we may write down the couplings of pions and rho mesons with $\Delta$ resonances as
\bea
{\cal L}_{\pi \Delta \Delta} = - \frac{\tilde g_A}{f_\pi}
\bar\psi^\sigma \gamma^\mu \gamma^5 \p_\mu \pi \psi_\sigma+
g_{\rho \Delta \Delta} \bar\psi^\sigma \gamma^\mu V_\mu
\psi_\sigma,
\eea
where the first term can be rewritten as
$2 ig_{\pi \Delta \Delta} \bar\psi^\sigma \gamma^5 \pi \psi_\sigma$, using the Goldberger-Treiman relation.

We now evaluate the above couplings from the bulk Lagrangian and
for this we need a proper identification of pions and rho mesons in
the bulk. As mentioned in the section two, we identify the pions
as the Nambu-Goldstone bosons of the broken chiral symmetry. We write
$X = v(z) e^{iP(x,z)}$ where $v(z) = \half M z+\half \sigma z^3$
and $P(x,z)$ is related to $A_{z}(x,z)$ by the relation~(\ref{ortho}). In the unitary gauge a linear
combination of $A_z$ and $P$ becomes the pion field when Kaluza-Klein reduced.
 We take without loss of generality
 \bea A_z(z,x) =
c_\pi f_0(z) \pi (x), \eea where $\pi(x)$ is identified as the  pion
field. The function $f_0(z)$ can be determined as in
\cite{Hong:2006ta, Hong:2007tf}.
Similarly,
the fifth component of vector gauge field gets decoupled in the
unitary gauge and the 4D vector components are
identified with the rho meson and its excited states. For the ground state we have
\bea V^\mu(z, x) = c_\rho g_0 (z)
\rho^\mu (x), \eea
where  $g_0(z) =
z J_1(m_{\rho}z)$, determined  from the wave equation for the vector gauge fields
in the bulk. The normalization constants $c_\pi$ and $c_\rho$ are
fixed by the requirement of canonical kinetic terms for pions and (4D)
vector gauge fields, respectively.

\subsection{$\pi$NN/ $\rho$NN couplings}

The pion (rho)-nucleon couplings can be estimated numerically from
the bulk interactions involving gauge fields, bulk scalars, and the nucleons.
The minimal gauge coupling and the Yukawa coupling with scalars are given as
\begin{equation}
{\cal L}_{\pi N N} = \bar{N}_1 \Gamma^z A_z N_1 -\bar{N}_2 \Gamma^z A_z N_2
- g (\bar{N}_1 X N_2 + \text{h.c.}).
\end{equation}
Since we identify the four dimensional vector components of the bulk
gauge fields with rho mesons, the rho meson couplings also arise from
the bulk gauge couplings,
\begin{equation}
{\cal L}_{\rho N N} =  \bar{N}_1 \Gamma^\mu V_\mu N_1 + \bar{N}_2
\Gamma^\mu V_\mu N_2.
\end{equation}
Using the Kaluza-Klein (KK) reduction of five dimensional spinors as
$N_{iL,R}(p,z) = \sum_n f^{(n)}_{iL,R}(p,z) \psi^{(n)}_{L,R}(p)$ ($i=1,2$), we
can write the four dimensional couplings for pions as
\begin{eqnarray}
g^{(0)nm}_{\pi NN} &= &-\int^{z_m}_0\! \!\frac{c_{\pi}dz}{2z^4}
\left[f_0\Big(f^{(n)*}_{1L} f^{(m)}_{1R} - f^{(n)*}_{2L}
f^{(m)}_{2R}\Big)\right.\nonumber\\
&& \qquad\qquad\qquad \left. -\frac{g z^2}{2 g^2_5
v(z)}\Big(\frac{f_0}{z}\Big)^\prime \Big(f^{(n)*}_{1L} f^{(m)}_{2R}
- f^{(n)*}_{2L} f^{(m)}_{1R}\Big)\right],
\end{eqnarray}
where
the prime denotes the derivative with respect to $z$.
(From now on we will absorb the normalization constants $c_{\pi}$ into $f_{0}$ and $c_{\rho}$ into $g_{0}$.) Similarly
the rho meson couplings can be written as
\begin{equation}
g^{(0)nm}_{\rho NN} =  \int^{z_m}_0 \frac{dz}{z^4} \,g_0(z)
\Big(f^{(n)*}_{1L}(z) f^{(m)}_{1L}(z) + f^{(n)*}_{2L}(z) f^{(m)}_{2L}(z)\Big).
\end{equation}

In addition to the minimal gauge interaction we should include the
following (parity-invariant) magnetic gauge couplings in
the bulk, as was done similarly in~\cite{Hong:2007tf},
\begin{eqnarray}
{\cal L}_{FNN} &=& i \kappa_1 \Big[\bar{N}_1 \Gamma^{MN} {(F_L)}_{MN} N_1 -
\bar{N}_2 \Gamma^{MN} {(F_R)}_{MN} N_2\Big] \nonumber\\
&& \, +\frac{i}{2}\kappa_2\Big[\bar{N}_1 X\Gamma^{MN} {(F_R)}_{MN} N_2 +
\bar{N}_2 X^{\dagger}\Gamma^{MN} {(F_L)}_{MN} N_1-\text{h.c.}\Big].\label{mag}
\end{eqnarray}
These terms contribute through the unknown coefficients $\kappa_1$ and $\kappa_{2}$.
Using the KK decomposition for the five dimensinal spinors as
before, the additional contribution to pion couplings can be written
as
\begin{eqnarray}
g^{(1)nm}_{\pi NN} &=& -(m^{(n)}_N+m^{(m)}_{N}) \int^{z_m}_0
\frac{dz}{z^3} f_0\left[\kappa_1 \Big(f^{(n)*}_{1L} f^{(m)}_{1L} +
f^{(n)*}_{2L}
f^{(m)}_{2L}\Big)\right.\nonumber\\
&&
\left.\qquad\qquad\qquad\qquad-\kappa_2 v(z) \Big(f^{(n)*}_{1L} f^{(m)}_{2L} -
f^{(n)*}_{2L} f^{(m)}_{1L}\Big)\right], \label{gpinn}
\end{eqnarray}
where $m^{(n)}_N$ denotes the nucleon mass of n-th excited state.
(Note that the second term in (\ref{gpinn}) vanishes
identically for the ground state nucleons or same excited states.) Similarly the rho meson
coupling has an extra contribution,
\begin{equation}
g^{(1)nm}_{\rho NN}\! =\! -2\!\int^{z_m}_0\!\! \frac{dz}{z^3}
g^{\prime}_0(z)\left[\kappa_1 \Big(f^{(n)*}_{1L} f^{(m)}_{1L}\! -\!
f^{(n)*}_{2L} f^{(m)}_{2L}\Big) +\kappa_2v(z)\Big(f^{(n)*}_{1L}
f^{(m)}_{2L} \!+\! f^{(n)*}_{2L} f^{(m)}_{1L}\Big)\right].
\end{equation}
The tensor coupling of rho mesons is determined by the magnetic gauge coupling~(\ref{mag}) as
\begin{equation}
f_{\rho}^{nm} \!= 4m_{N}\!\!\int^{z_m}_0\!\! \frac{dz}{z^3}
g_0(z)\left[\kappa_1 \Big(f^{(n)*}_{1L} f^{(m)}_{1R}\! -\!
f^{(n)*}_{2L} f^{(m)}_{2R}\Big) + \kappa_2v(z)\Big(f^{(n)*}_{1L}
f^{(m)}_{2R} \!+\! f^{(n)*}_{2L} f^{(m)}_{1R}\Big)\right].
\end{equation}

For the rho and $\pi$ meson couplings we have two unknown parameters
$\kappa_{1}$ and $\kappa_{2}$. Fitting the couplings $g_{\pi
NN}=13.5$, $g_{\rho NN}=-8.6$ of the ground state nucleons, we get
$\kappa_{1}=-0.98$ and $\kappa_{2}=1.25$, which then determine the
tensor coupling $f_{\rho}=-19.6$ for the ground state nucleons,
larger than the value, quoted in~\cite{Riska:2000gd}. We
predict all other couplings of rho and $\pi$ mesons with (excited)
nucleons. Some are shown in Table~\ref{t1}  and~\ref{t2}  for two
different sets of values for the fitting parameters $\kappa_{1}$ and
$\kappa_{2}$, respectively. The numerical values are compared with
the calculated ones from the chiral quark model (quoted from the
reference~\cite{Riska:2000gd}), denoted in the  brackets.
\begin{table}[htb]
\begin{tabular}{|c|c|c|c|c|c|}
\hline
$g_{\pi NN}$& 13.5 (13.5) & $g_{\rho NN}$ & -8.6 (2.8) & $f_{\rho}$ & -19.6 \\
\hline
 $g^{(1440)}_{\pi NN^*}$ & -20.19 (0.26 $g_{\pi NN}$) & $g^{(1440)}_{\rho NN^*}$
 & 25.94 (-3.1) & $f^{(1440)}_{\rho}$ & -18.72 \\
\hline
 $g^{(1535)}_{\pi NN^*}$ & -7.12 (0.49 $g_{\pi NN}$) & $g^{(1535)}_{\rho NN^*}$
 & 4.45 (4.8) & $f^{(1535)}_{\rho}$ & -19.01 \\
\hline
\end{tabular}
\caption{Numerical result for $\kappa_1 = -0.98$ and $\kappa_2 =
1.25$ and we have used $\sigma = 0.85/z^3_m - m_q/z^2_m$ with
$z^{-1}_m = 205 \text{MeV} $.}\label{t1}
\end{table}

\begin{table}[htb]
\begin{tabular}{|c|c|c|c|c|c|}
\hline
$g_{\pi NN}$& 11.54 (13.5) & $g_{\rho NN}$ & -3.42 (2.8) & $f_{\rho}$ & -21.10 \\
\hline
 $g^{(1440)}_{\pi NN^*}$ & -13.96 (0.26 $g_{\pi NN}$) & $g^{(1440)}_{\rho NN^*}$ & 19.8 (-3.1)
 & $f^{(1440)}_{\rho}$ & -20.18 \\
\hline
 $g^{(1535)}_{\pi NN^*}$ & -5.45 (0.49 $g_{\pi NN}$) & $g^{(1535)}_{\rho NN^*}$   & 1.59 (4.8)
 & $f^{(1535)}_{\rho}$ & -20.49 \\
\hline
\end{tabular}
\caption{Numerical result for $\kappa_1 = -0.78$ and $\kappa_2 = 0.5
$ and we have used $\sigma = 0.85/z^3_m- m_q/z^2_m$ with $z^{-1}_m =
205 \text{MeV} $.}\label{t2}
\end{table}

\subsection{$\pi \Delta \Delta/\rho \Delta \Delta$ couplings}

The $\pi$-$\Delta$ and rho-$\Delta$ couplings can be estimated numerically from
the bulk interactions involving gauge fields and $\Delta$
resonances,
\begin{equation}
{\cal L}_{\pi \Delta \Delta} =
\bar\Psi_1^{\mu} \Gamma^z A_z \Psi_{1\mu} -\bar\Psi_2^{\mu} \Gamma^z
A_z \Psi_{2\mu} - g_{3/2} (\bar\Psi_1^{\mu} X^3 \Psi_{2\mu} + \text{h.c.})\,.
\end{equation}
Using the KK reduction for the five dimensional spinors
$\Psi^\sigma _{L,R}(p,z) = \sum_n F^{(n)}_{L,R}(p,z) \psi^{(n)\sigma}_{L,R
} (p)$, we can write the four dimensional
couplings for pions as
\begin{eqnarray}
g^{(0)nm}_{\pi \Delta \Delta} &=& \! -\int^{z_m}_0 \frac{dz}{2z^2}
\left[f_0\Big(F^{(n)*}_{1L} F^{(m)}_{1R} - F^{(n)*}_{2L}
F^{(m)}_{2R}\Big)\right.\nonumber\\
&&\qquad\qquad~~\left. -\frac{3g_{3/2} z^2v(z)}{2
g^2_5}\left(\frac{f_0}{z}\right)^\prime \Big(F^{(n)*}_{1L}
F^{(m)}_{2R} - F^{(n)*}_{2L} F^{(m)}_{1R}\Big)\right].
\end{eqnarray}
Similarly the rho meson couplings arise from the bulk gauge
couplings,
\begin{equation}
{\cal L}_{\rho \Delta \Delta} = \bar\Psi_1^{\nu} \Gamma^\mu V_\mu
\Psi_{1\nu} + \bar\Psi_2^{\nu} \Gamma^\mu V_\mu
\Psi_{2\nu},
\end{equation}
which can be written as
\begin{equation}
g^{(0)nm}_{\rho \Delta \Delta} = \int^{z_m}_0 \frac{dz}{z^2} g_0
\Big(F^{(n)*}_{1L} F^{(m)}_{1L} + F^{(n)*}_{2L} F^{(m)}_{2L}\Big).
\end{equation}

The numerical values from the above coupling are too small to
account for the experimental values. The additional contributions to
$\pi$-$\Delta$ and  rho-$\Delta$ couplings can arise from the following magnetic
type of couplings in the bulk, similarly to the couplings with (excited) nucleons,
\begin{eqnarray}
{\cal L}_{F\Delta \Delta} &=& i \kappa_3 \Big[\bar\Psi_1^M \Gamma^{NP}
{(F_L)}_{NP} \Psi_{1M} - \bar\Psi_2^{M} \Gamma^{NP} {(F_R)}_{NP}
\Psi_{2M}\Big] \nonumber\\
&& \, +\frac{i}{2} \kappa_4 \Big[\bar\Psi_1^M X^3\Gamma^{NP} {(F_R)}_{NP}
\Psi_{2M} + \bar\Psi_2^{M} (X^{\dagger})^3\Gamma^{NP} {(F_L)}_{NP}
\Psi_{1M}-\text{h.c.}\Big].
\end{eqnarray}

Using the KK mode decomposition for the spinors as before, the
additional contribution for pion couplings can be written as,
\begin{eqnarray}
g^{(1)nm}_{\pi \Delta \Delta} &=&
-(m^{(n)}_{\Delta}+m^{(m)}_{\Delta})\int^{z_m}_0 \frac{dz}{z} f_0
\left[\kappa_3\Big(F^{(n)*}_{1L} F^{(m)}_{1L} + F^{(n)*}_{2L}
F^{(m)}_{2L}\Big)\right.\nonumber\\
&&\qquad\qquad\qquad\qquad\qquad~~ \left. -\kappa_4 (v(z))^3\Big(F^{(n)*}_{1L} F^{(m)}_{2L}
- F^{(n)*}_{2L} F^{(m)}_{1L}\Big)\right],
\end{eqnarray}
and similarly for the rho meson couplings as,
\begin{equation}
g^{(1)nm}_{\rho \Delta \Delta} =\! -2\!\int^{z_m}_0 \!\frac{dz}{z}
g_0^{\prime} \left[\kappa_3\Big(\!F^{(n)*}_{1L} F^{(m)}_{1L} \!-\!
F^{(n)*}_{2L} F^{(m)}_{2L}\!\Big)\! +\! \kappa_4
(v(z))^3\Big(\!F^{(n)*}_{1L} F^{(m)}_{2L} \!+\! F^{(n)*}_{2L}
F^{(m)}_{1L}\!\Big)\right].
\end{equation}

We fix the value of $\kappa_3$ and $\kappa_4$ by fitting the
coupling $g_{\pi\Delta\Delta}=20$ and $g_{\rho\Delta\Delta}=10.9$
for the ground state $\Delta$-baryons, respectively. We present the
$\pi$-$\Delta$ and rho-$\Delta$ couplings in
 Table~\ref{t3}.

\begin{table}[htb]
\begin{tabular}{|c|c|c|c|}
\hline
$g_{\pi \Delta\Delta}$ & 20  & $g_{\rho  \Delta\Delta}$ & 10.9  \\
\hline
 $g^{(1700)}_{\pi  \Delta\Delta^*}$ & 43.78 & $g^{(1700)}_{\rho  \Delta\Delta^*}$ & 37.40 \\
\hline
$g^{(1920)}_{\pi  \Delta\Delta^*}$ & -79.94 &$g^{(1920)}_{\rho  \Delta\Delta^*}$ & 45.35 \\
\hline
\end{tabular}
\caption{Numerical result for $\kappa_3 = 0.07 $ and $\kappa_4 =
11.32$ and we have used $\sigma = 0.85/z^3_m - m_q/z^2_m$ with
$z^{-1}_m = 205\text{ MeV} $.} \label{t3}
\end{table}

\subsection{$\pi N \Delta/\rho N \Delta$ couplings}

The transition couplings of pions is determined from the gauge
invariant couplings of gauge fields with nucleons and $\Delta$
resonances in the bulk. The Lagrangian is given by
\begin{eqnarray}\label{Sfnd}
{\cal
L}_{FN\Delta}&=&\left[\alpha_1\Big(\bar{\Psi}_1^M\Gamma^N(F_L)_{MN}N_1-(1\leftrightarrow
2 \,\, \&\,\, L\leftrightarrow R)\Big)\right. \nonumber\\
&&\left.~
+i\alpha_2\Big((\partial^M\bar{\Psi}_1^N)(F_L)_{MN}N_1+(1\leftrightarrow
2 \,\, \&\,\, L\leftrightarrow R)\Big)\right. \nonumber\\
&&\left.~
+i\alpha_3\Big(\bar{\Psi}_1^M(\partial^N(F_L)_{MN})N_1+(1\leftrightarrow
2 \,\, \&\,\, L\leftrightarrow R)\Big)\right. \nonumber\\
&&\left.~
+\beta_1\Big(\bar{\Psi}_1^M\Gamma^{NP}(\tilde{F}_L)_{MNP}N_1-(1\leftrightarrow
2 \,\, \&\,\, L\leftrightarrow R)\Big)\right. \nonumber\\
&&\left.~
+i\beta_2\Big((\partial^M\bar{\Psi}_1^N)\Gamma^P(\tilde{F}_L)_{MNP}N_1+(1\leftrightarrow
2 \,\, \&\,\, L\leftrightarrow R)\Big)\right.
\nonumber\\
&&\left.~
+\tilde{\alpha}_1\Big(\bar{\Psi}_1^M\Gamma^N(F_L)_{MN}XN_2+(1\leftrightarrow
2 \,\, \&\,\, L\leftrightarrow R)\Big)\right. \nonumber\\
&&\left.~
+i\tilde{\alpha}_2\Big((\partial^M\bar{\Psi}_1^N)(F_L)_{MN}XN_2-(1\leftrightarrow
2 \,\, \&\,\, L\leftrightarrow R)\Big)\right. \nonumber\\
&&\left.~
+i\tilde{\alpha}_3\Big(\bar{\Psi}_1^M(\partial^N(F_L)_{MN})XN_2-(1\leftrightarrow
2 \,\, \&\,\, L\leftrightarrow R)\Big)\right. \nonumber\\
&&\left.~
+\tilde{\beta}_1\Big(\bar{\Psi}_1^M\Gamma^{NP}(\tilde{F}_L)_{MNP}XN_2+(1\leftrightarrow
2 \,\, \&\,\, L\leftrightarrow R)\Big)\right. \nonumber\\
&&\left.~
+i\tilde{\beta}_2\Big((\partial^M\bar{\Psi}_1^N)\Gamma^P(\tilde{F}_L)_{MNP}XN_2-(1\leftrightarrow
2 \,\, \&\,\, L\leftrightarrow R)\Big) +{\rm{h.c.}} \right],
\end{eqnarray}
where $\alpha$'s and $\beta$'s are unknown parameters. The same term
also contributes to the four dimensional rho meson couplings. Using
the KK reduction of five dimensional spinors as $\Psi_{iL,R}(p,z) =
\sum_n f^{(n)}_{iL,R}(p,z) \psi^{(n)}_{L,R} (p)$  for nucleons and
$\Psi_{L,R}(p,z) = \sum_n F^{(n)}_{L,R}(p,z) \psi^{(n)}_{L,R} (p)$ 
for $\Delta$ resonances, we can write the pion-nucleon-$\Delta$
couplings as,
\begin{eqnarray}
g^{nm}_{\pi N \Delta} &=& -f_{\pi}\int^{z_m}_0 dz
\left[\frac{f_0}{z^2}\Big(\kappa\big(F^{(n)*}_{1L} f^{(m)}_{1R} +
F^{(n)*}_{2L} f^{(m)}_{2R}\big) + \tilde{\kappa}
v(z)\big(F^{(n)*}_{1L} f^{(m)}_{2R} - F^{(n)*}_{2L} f^{(m)}_{1R}\big)\Big)\right. \nonumber\\
&&\left. +\frac{f_0}{z}\Big(\alpha_2\big((\partial_zF^{(n)*}_{1L})
f^{(m)}_{1R} - (\partial_zF^{(n)*}_{2L}) f^{(m)}_{2R}\big)
+\tilde{\alpha}_2 v(z)\big((\partial_zF^{(n)*}_{1L})
f^{(m)}_{2R} + (\partial_zF^{(n)*}_{2L}) f^{(m)}_{1R}\big)\Big)\right. \nonumber\\
&&\left. -\frac{f_0^{\prime}}{z}\Big(\alpha_3\big(F^{(n)*}_{1L}
f^{(m)}_{1R} - F^{(n)*}_{2L} f^{(m)}_{2R}\big) + \tilde{\alpha}_3
v(z)\big(F^{(n)*}_{1L} f^{(m)}_{2R} + F^{(n)*}_{2L}
f^{(m)}_{1R}\big)\Big)\right. \nonumber\\
&&\left. +2m_{\Delta}\frac{f_0}{z}\Big(\beta_2\big(F^{(n)*}_{1L}
f^{(m)}_{1L} - F^{(n)*}_{2L} f^{(m)}_{2L}\big) + \tilde{\beta}_2
v(z)\big(F^{(n)*}_{1L} f^{(m)}_{2L} + F^{(n)*}_{2L}
f^{(m)}_{1L}\big)\Big)\right],
\end{eqnarray}
and similarly the rho-nucleon-$\Delta$ couplings as,
\begin{eqnarray}
g^{nm}_{\rho N \Delta} &=& \int^{z_m}_0 dz
\left[\frac{g_0^{\prime}}{z^2}\Big(\kappa \big(F^{(n)*}_{1L}
f^{(m)}_{1R} - F^{(n)*}_{2L} f^{(m)}_{2R}\big) + \tilde{\kappa}
v(z)\big(F^{(n)*}_{1L} f^{(m)}_{2R} + F^{(n)*}_{2L} f^{(m)}_{1R}\big)\Big)\right. \nonumber\\
&&\left.
+\frac{g_0^{\prime}}{z}\Big(\alpha_2\big((\partial_zF^{(n)*}_{1L})
f^{(m)}_{1R} + (\partial_zF^{(n)*}_{2L}) f^{(m)}_{2R}\big)
+\tilde{\alpha}_2 v(z)\big((\partial_zF^{(n)*}_{1L})
f^{(m)}_{2R} - (\partial_zF^{(n)*}_{2L}) f^{(m)}_{1R}\big)\Big)\right. \nonumber\\
&&\left.
-\frac{g_0^{\prime\prime}}{z}\Big(\alpha_3\big(F^{(n)*}_{1L}
f^{(m)}_{1R} + F^{(n)*}_{2L} f^{(m)}_{2R}\big) + \tilde{\alpha}_3
v(z)\big(F^{(n)*}_{1L} f^{(m)}_{2R} - F^{(n)*}_{2L}
f^{(m)}_{1R}\big)\Big)\right. \nonumber\\
&&\left.
-2m_{\Delta}\frac{g_0^{\prime}}{z}\Big(\beta_2\big(F^{(n)*}_{1L}
f^{(m)}_{1L} + F^{(n)*}_{2L} f^{(m)}_{2L}\big) + \tilde{\beta}_2
v(z)\big(F^{(n)*}_{1L} f^{(m)}_{2L} - F^{(n)*}_{2L}
f^{(m)}_{1L}\big)\Big)\right],
\end{eqnarray}
where $\kappa=\alpha_1-4\beta_1$ and
$\tilde{\kappa}=\tilde{\alpha}_1-4\tilde{\beta}_1$.

The nucleon-$\Delta$ couplings with pions or rho-mesons have eight unknown parameters ($\kappa$, $\tilde\kappa_{2}$,
$\alpha_{2,3}$, $\tilde\alpha_{2,3}$, $\beta_{2}$, $\tilde\beta_{2}$).
We fix those parameters in section~\ref{ff}, where we calculate the nucleon-$\Delta$ transition form-factors, and we get $g_{\pi N \Delta}= 20.93 (1.55 g_{\pi NN})$  and  $g_{\rho N \Delta}=8.7 (8.7)$.   (The values in the bracket are quoted from
the reference \cite{Riska:2000gd}). In principle we can calculate the transition couplings for the excited $\Delta$ resonances, but our hard-wall model does not seem to work.

\section{Nucleon to $\Delta$ Transition Form Factors}
\label{ff}
In this section we evaluate the  nucleon to $\Delta$  transition form-factors
in AdS/QCD and present the numerical values for the form factors.
The nucleon-$\Delta$ electromagnetic and axial transition form-factors
in four dimensions are extracted from the matrix elements of
the vector and axial vector currents between $\Delta$ and nucleon
states,
\begin{equation}\label{EMmat}
\langle\Delta (p')|J_{\mu}^{EM}|N(p)\rangle =
\bar{u}^{\sigma}(p'){\cal{O}}^{(EM)}_{\sigma\mu}(p,p')u(p),
\end{equation}
\begin{equation}\label{Amat}
\langle\Delta (p')|A_{\mu}^{3}|N(p)\rangle =
\bar{u}^{\sigma}(p'){\cal{O}}^{(A)}_{\sigma\mu}(p,p')u(p),
\end{equation}
where $u^{\sigma}(p')$ and $u(p)$ are the Rarita-Schwinger and
nucleon spinors for $\Delta$  and nucleon of momentum of momentum
$p'$ and $p$, respectively.

By the Lorentz invariance
and the current conservation we expand the operators ${\cal{O}}^{(EM)}_{\sigma\mu}$~\cite{Jones:1972ky} and
${\cal{O}}^{(A)}_{\sigma\mu}$~\cite{Alexandrou:2007zz}, assuming the CP invariance, respectively as
\begin{equation}\label{EMff1}
{\cal{O}}^{(EM)}_{\sigma\mu}\!=\!
\left[G_1(q^2)(q_{\sigma}\gamma_{\mu}\!-\!q\cdot\gamma\eta_{\sigma\mu})\!+\!G_2(q^2)(q_{\sigma}P_{\mu}\!-\!q\cdot
P\eta_{\sigma\mu})\!+\!G_3(q^2)(q_{\sigma}q_{\mu}\!-\!q^2\eta_{\sigma\mu})
\right]\gamma^5,
\end{equation}
\begin{equation}\label{Aff}
{\cal{O}}^{(A)}_{\sigma\mu}\!=\!
\frac{C^A_3(q^2)}{m_N}(q\cdot\gamma\eta_{\sigma\mu}\!-\!q_{\sigma}\gamma_{\mu})\!+\!\frac{C^A_4(q^2)}{m_N^2}(p'\cdot
q\eta_{\sigma\mu}\!-\!q_{\sigma}p'_{\mu})\!+\!C^A_5(q^2)\eta_{\sigma\mu}\!+\!\frac{C^A_6(q^2)}{m_N^2}q_{\sigma}q_{\mu},
\end{equation}
where $q=p'-p$ and $P=(p'+p)/2$. All $C^A_i$'s are
dimensionless, but $G_1$  has the dimension of mass inverse and both $G_2$ and $G_3$ have the dimension of
mass-inverse squared. The
pion-nucleon-$\Delta$ form factor, $G_{\pi N \Delta}$, is related to
$C^A_5$ by the Goldberger-Treiman relation \cite{Alexandrou:2007zz}
\begin{equation}
G_{\pi N \Delta}(q^2) = \frac{2m_N}{f_{\pi}}C^A_5(q^2),
\end{equation}
and $\pi N \Delta$ coupling constant is given by $g_{\pi N \Delta} =
G_{\pi N \Delta}(0)$.

The vertex operater for the electromagnetic matrix elements eq.~(\ref{EMff1}) can be also expressed in terms of the three Sachs form
factors~\cite{Alexandrou:2007dt, Jones:1972ky} as follow
\begin{equation}\label{EMff2}
{\cal{O}}^{(EM)}_{\sigma\mu}(p',p)=G_{M1}(q^2)K^{M1}_{\sigma\mu}+G_{E2}(q^2)K^{M1}_{\sigma\mu}
+G_{C2}(q^2)K^{M1}_{\sigma\mu},
\end{equation}
where $G_{M1}$, $G_{E2}$ and $G_{C2}$ are magnetic dipole, electric
quadrupole and Coulomb quadrupole form factors, respectively and
\begin{eqnarray}
K^{M1}_{\sigma\mu}&=&-\frac{3}{(m_{\Delta}+m_N)^2-q^2}\frac{m_{\Delta}+m_N}{2m_N}
i\epsilon_{\sigma\mu\alpha\beta}p^{\alpha}p'^{\beta}, \\
K^{E2}_{\sigma\mu}&=&-K^{M1}_{\sigma\mu}+6\Omega^{-1}(q^2)\frac{m_{\Delta}+m_N}{2m_N}
2i\gamma^5\epsilon_{\sigma\lambda\alpha\beta}p^{\alpha}p'^{\beta}{\epsilon_{\mu}}^{\lambda\gamma\delta}p_{\gamma}p'_{\delta},
\\
K^{C2}_{\sigma\mu}&=&-6\Omega^{-1}(q^2)\frac{m_{\Delta}+m_N}{2m_N}i\gamma^5q_{\sigma}(q^2(p+p')_{\mu}-q\cdot
(p+p')q_{\mu},
\end{eqnarray}
where
\begin{equation}
\Omega(q^2)=\left[(m_{\Delta}+m_N)^2-q^2\right]\left[(m_{\Delta}-m_N)^2-q^2\right].
\end{equation}
The $G_{M1}$, $G_{E2}$ and $G_{C2}$ are related to  $G_i$'s
\cite{Jones:1972ky} as, with $\hat m\equiv m_{N}/(m_{\Delta}+m_{N})$,
\begin{eqnarray}
G_{M1}(q^2)\!\!\!&=&\!\!\!\left[\frac{(3m_{\Delta}\!+\!m_N)(m_{\Delta}\!+\!m_N)-q^2}{m_{\Delta}}G_1(q^2)\!
+\!(m_{\Delta}^2\!-\!m_N^2)G_2(q^2)+2q^2G_3(q^2)\right]\frac{\hat m}{3}, \\
G_{E2}(q^2)\!\!\!&=&\!\!\!\left[\frac{(m_{\Delta}^2-m_N^2)+q^2}{m_{\Delta}}G_1(q^2)
+(m_{\Delta}^2-m_N^2)G_2(q^2)+2q^2G_3(q^2)\right]\frac{\hat m}{3}, \\
G_{M1}(q^2)\!\!\!&=&\!\!\!\left[2m_{\Delta}G_1(q^2)
+\frac{1}{2}(3m_{\Delta}^2+m_N^2-q^2)G_2(q^2)+(m_{\Delta}^2-m_N^2+q^2)G_3(q^2)\right]\frac{2{\hat m}}{3}.
\end{eqnarray}
The ratios of the electric and Coulomb quadrupole amplitudes to the
magnetic dipole amplitude, $R_{EM}$ (EMR) and $R_{SM}$ (CMR), are
defined  as
\begin{equation}
R_{EM}=-\frac{G_{E2}(q^2)}{G_{M1}(q^2)}, \qquad\qquad
R_{SM}=-\frac{|\vec{q}|}{2m_{\Delta}}\frac{G_{C2}(q^2)}{G_{M1}(q^2)}.
\end{equation}

The 5D action for the nucleon to $\Delta$ coupling, corresponding
to the matrix elements eqs. (\ref{EMmat}) and (\ref{Amat}), is given
in eq.~(\ref{Sfnd}). By the AdS/CFT
correspondence, the matrix element is given from the bulk action by taking the normalizable
modes for the nucleon and $\Delta$, and non-normalizable modes for (external) vector and axial vector gauge
fields. Taking the axial gauge, where $V_z=0=A_z$,  we Fourier-transform the vector and axial vector gauge fields  as
\begin{eqnarray}
V_{\mu}(x,z)=\int_q V_{\mu}(q)V(q,z)e^{-iq\cdot x}, \qquad
A_{\mu}(x,z)=\int_q A_{\mu}(q)A(q,z)e^{-iq\cdot x},
\end{eqnarray}
with boundary conditions, $V(q,\epsilon)=1$ at UV
$(z=\epsilon)$ and $\partial_z V(q,z_m)=0$ at IR $(z=z_{m})$,
and similarly for $A(q,z)$. The 5D wave functions of the
gauge fields, $V(q,z)$ and $A(q,z)$, are determined by solving the
equations of motions. We decompose  Rarita-Schinger fields into chirality basis  as
\begin{equation}
\Psi^{\mu}_{i}(z,x)=\int_{p}
\left[F_{iL}(p,z)u^{\mu}_L(p)+F_{iR}(p,z)u^{\mu}_R(p)\right]e^{-ip\cdot x},\quad (i=1,2)
\end{equation}
and similarly for the nucleon fields but with $f(p,z)$ and $u(p)$
instead of $F(p,z)$ and $u^{\mu}(p)$.

By matching the operators from the 5D Lagrangian~(\ref{Sfnd})
with eq.'s~(\ref{EMff1}) and (\ref{Aff}), we easily read off the nucleon to
$\Delta$ transition form-factors and get
\begin{eqnarray}
G_1(Q^2)&=&\int
dz\left[\frac{V(q,z)}{z^2}\Big(\kappa\big(F^{\ast}_{1L}(p^{\prime},z)f_{1L}(p,z)-F^{\ast}_{2L}(p^{\prime},z)f_{2L}(p,z)\big)\Big.\right.
\nonumber\\
&&\left.\Big.\qquad\qquad\qquad
+\tilde{\kappa}v(z)\big(F^{\ast}_{1L}(p^{\prime},z)f_{2L}(p,z)+F^{\ast}_{2L}(p^{\prime},z)f_{1L}(p,z)\big)\Big)
\right. \nonumber\\
&&\left.\qquad\quad
+\frac{2m_{\Delta}V(q,z)}{z}\Big(\beta_2\big(F^{\ast}_{1L}(p^{\prime},z)f_{1R}(p,z)+F^{\ast}_{2L}(p^{\prime},z)f_{2R}(p,z)\big)\Big.\right.
\nonumber\\
&&\left.\Big.\qquad\qquad\qquad\qquad\quad
+\tilde{\beta}_2v(z)\big(F^{\ast}_{1L}(p^{\prime},z)f_{2R}(p,z)-F^{\ast}_{2L}(p^{\prime},z)f_{1R}(p,z)\big)\Big)
\right. \nonumber\\
&&\left.\qquad\quad
+\frac{2V(q,z)}{z}\Big(\beta_2\big((\partial_zF^{\ast}_{1L}(p^{\prime},z))f_{1L}(p,z)+(\partial_zF^{\ast}_{2L}(p^{\prime},z))f_{2L}(p,z)\big)\Big.\right.
\nonumber\\
&&\left.\Big.\qquad\qquad
+\tilde{\beta}_2v(z)\big((\partial_zF^{\ast}_{1L}(p^{\prime},z))f_{2L}(p,z)-(\partial_zF^{\ast}_{2L}(p^{\prime},z))f_{1L}(p,z)\big)\Big)\right],
\end{eqnarray}
\begin{eqnarray}
G_2(Q^2)&=&-\int dz\left[\frac{V(q,z)}{z}\Big((
\alpha_2+2\beta_2)\big(F^{\ast}_{1L}(p^{\prime},z)f_{1R}(p,z)+F^{\ast}_{2L}(p^{\prime},z)f_{2R}(p,z)\big)\Big.\right.
\nonumber\\
&&\left.\Big.\qquad\qquad
+(\tilde{\alpha}_2+2\tilde{\beta}_2)v(z)\big(F^{\ast}_{1L}(p^{\prime},z)f_{2R}(p,z)-F^{\ast}_{2L}(p^{\prime},z)f_{1R}(p,z)\big)\Big)
\right],
\end{eqnarray}
\begin{eqnarray}
G_3(Q^2)&=&-\int
dz\left[\frac{V(q,z)}{2z}\Big((\alpha_2+2\beta_2+2\alpha_3)\big(F^{\ast}_{1L}(p^{\prime},z)f_{1R}(p,z)+F^{\ast}_{2L}(p^{\prime},z)f_{2R}(p,z)\big)\Big.\right.
\nonumber\\
&&\left.\Big.\qquad
+(\tilde{\alpha}_2+2\tilde{\beta}_2+2\tilde{\beta}_3)v(z)\big(F^{\ast}_{1L}(p^{\prime},z)f_{2R}(p,z)-F^{\ast}_{2L}(p^{\prime},z)f_{1R}(p,z)\big)\Big)
\right]
\end{eqnarray}
for vectors, and
\begin{eqnarray}
C^A_3(Q^2)&=&-m_N\int
dz\left[\frac{A(q,z)}{z^2}\Big(\kappa\big(F^{\ast}_{1L}(p^{\prime},z)f_{1L}(p,z)+F^{\ast}_{2L}(p^{\prime},z)f_{2L}(p,z)\big)\Big.\right.
\nonumber\\
&&\left.\Big.\qquad\qquad\qquad\qquad
+\tilde{\kappa}v(z)\big(F^{\ast}_{1L}(p^{\prime},z)f_{2L}(p,z)-F^{\ast}_{2L}(p^{\prime},z)f_{1L}(p,z)\big)\Big)
\right. \nonumber\\
&&\left.\qquad\qquad\quad
-\frac{2m_{\Delta}A(q,z)}{z}\Big(\beta_2\big(F^{\ast}_{1L}(p^{\prime},z)f_{1R}(p,z)-F^{\ast}_{2L}(p^{\prime},z)f_{2R}(p,z)\big)\Big.\right.
\nonumber\\
&&\left.\Big.\qquad\qquad\qquad\qquad\qquad\quad
+\tilde{\beta}_2v(z)\big(F^{\ast}_{1L}(p^{\prime},z)f_{2R}(p,z)+F^{\ast}_{2L}(p^{\prime},z)f_{1R}(p,z)\big)\Big)
\right. \nonumber\\
&&\left.\qquad\qquad\quad
+\frac{2A(q,z)}{z}\Big(\beta_2\big((\partial_zF^{\ast}_{1L}(p^{\prime},z))f_{1L}(p,z)-(\partial_zF^{\ast}_{2L}(p^{\prime},z))f_{2L}(p,z)\big)\Big.\right.
\nonumber\\
&&\left.\Big.\qquad\qquad~~
+\tilde{\beta}_2v(z)\big((\partial_zF^{\ast}_{1L}(p^{\prime},z))f_{2L}(p,z)+(\partial_zF^{\ast}_{2L}(p^{\prime},z))f_{1L}(p,z)\big)\Big)\right],
\end{eqnarray}
\begin{eqnarray}
C^A_4(Q^2)&=&-m^2_N\int
dz\left[\frac{A(q,z)}{z}\Big((\alpha_2+2\beta_2)\big(F^{\ast}_{1L}(p^{\prime},z)f_{1R}(p,z)-F^{\ast}_{2L}(p^{\prime},z)f_{2R}(p,z)\big)\Big.\right.
\nonumber\\
&&\left.\Big.\qquad~~
+(\tilde{\alpha}_2+2\tilde{\beta}_2)v(z)\big(F^{\ast}_{1L}(p^{\prime},z)f_{2R}(p,z)+F^{\ast}_{2L}(p^{\prime},z)f_{1R}(p,z)\big)\Big)
\right],
\end{eqnarray}
\begin{eqnarray}
C^A_5(Q^2)&=&\int
dz\left[\frac{\partial_zA(q,z)}{z^2}\Big(\kappa\big(F^{\ast}_{1L}(p^{\prime},z)f_{1R}(p,z)+F^{\ast}_{2L}(p^{\prime},z)f_{2R}(p,z)\big)\Big.\right.
\nonumber\\
&&\left.\Big.\qquad\qquad\qquad\quad
+\tilde{\kappa}v(z)\big(F^{\ast}_{1L}(p^{\prime},z)f_{2R}(p,z)-F^{\ast}_{2L}(p^{\prime},z)f_{1R}(p,z)\big)\Big)
\right. \nonumber\\
&&\left.\qquad\quad
+\frac{\partial_zA(q,z)}{z}\Big(\alpha_2\big((\partial_zF^{\ast}_{1L}(p^{\prime},z))f_{1R}(p,z)-(\partial_zF^{\ast}_{2L}(p^{\prime},z))f_{2R}(p,z)\big)\Big.\right.
\nonumber\\
&&\left.\Big.\qquad\qquad\qquad\qquad
+\tilde{\alpha}_2v(z)\big((\partial_zF^{\ast}_{1L}(p^{\prime},z))f_{2R}(p,z)+(\partial_zF^{\ast}_{2L}(p^{\prime},z))f_{1R}(p,z)\big)\Big)
\right. \nonumber\\
&&\left.\qquad\quad
-\frac{\partial^2_zA(q,z)}{z}\Big(\alpha_3\big(F^{\ast}_{1L}(p^{\prime},z)f_{1R}(p,z)-F^{\ast}_{2L}(p^{\prime},z)f_{2R}(p,z)\big)\Big.\right.
\nonumber\\
&&\left.\Big.\qquad\qquad\qquad\qquad
+\tilde{\alpha}_3v(z)\big(F^{\ast}_{1L}(p^{\prime},z)f_{2R}(p,z)+F^{\ast}_{2L}(p^{\prime},z)f_{1R}(p,z)\big)\Big)\right. \nonumber\\
&&\left.\qquad\quad
+\frac{2m_{\Delta}\partial_zA(q,z)}{z}\Big(\beta_2\big(F^{\ast}_{1L}(p^{\prime},z)f_{1L}(p,z)-F^{\ast}_{2L}(p^{\prime},z)f_{2L}(p,z)\big)\Big.\right.
\nonumber\\
&&\left.\Big.\qquad\qquad\qquad
+\tilde{\beta}_2v(z)\big(F^{\ast}_{1L}(p^{\prime},z)f_{2L}(p,z)+F^{\ast}_{2L}(p^{\prime},z)f_{1L}(p,z)\big)\Big)\right],
\end{eqnarray}
\begin{eqnarray}
C^A_6(Q^2)&=&m^2_N\int
dz\frac{A(q,z)}{z}\Big[\alpha_3\big(F^{\ast}_{1L}(p^{\prime},z)f_{1R}(p,z)-F^{\ast}_{2L}(p^{\prime},z)f_{2R}(p,z)\big)\Big.
\nonumber\\
&&\Big.\qquad\qquad
+\tilde{\alpha}_3v(z)\big(F^{\ast}_{1L}(p^{\prime},z)f_{2R}(p,z)+F^{\ast}_{2L}(p^{\prime},z)f_{1R}(p,z)\big)\Big],
\end{eqnarray}
for axial vector, where $\kappa=\alpha_1-4\beta_1$,
$\tilde{\kappa}=\tilde{\alpha}_1-4\tilde{\beta}_1$ and $Q^2=-q^2$.

\subsection{Numerical Results}
In order to perform the numerical calculation we set three
parameters, $z_m$, $m_q$, and $\sigma$, to $z_m=(205
{\rm{Mev}})^{-1}$, $m_q=0.0023$ and $\sigma=0.85/z^3_m - m_q/z^2_m$.
We also determine the unknown parameters
\begin{eqnarray}
\kappa&=&-179,~\tilde\kappa=-733,~\alpha_{2}=18,~\tilde\alpha_{2}=136,~ \nn\\
\alpha_{3}&=&51,~\tilde\alpha_{3}=-148,~\beta_{2}=-9,~\tilde\beta_{2}=-68\,,
\end{eqnarray}
by setting the numerical
values of our $\pi N \Delta$, $\rho N \Delta$ coupling constants to
be the same as the experimental values and $N-\Delta$ form factors
to be the same as the lattice QCD values at a fixed value of $Q^2$,
i.e.,
\begin{eqnarray}
g_{\pi N \Delta}=1.55 g_{\pi N \Delta}=20.93, \quad
g_{\rho N \Delta}=8.7 \qquad\qquad    &&{\text{\cite{Riska:2000gd}}}, \nonumber\\
C^A_3(0)=0, \quad  C^A_4(0)=0, \quad  C^A_5(0)=\frac{f_{\pi}}{2m_{N}}g_{\pi N\Delta}=1.11, \quad
C^A_6(0.15)=2.266 &&   {\text{\cite{Alexandrou:2007zz}}},
\nonumber\\
G_1(0)=2.38 \quad  G_2(0)=-0.56 \qquad\qquad &&
{\text{\cite{Jones:1972ky}}}.
\end{eqnarray}

\newpage
The numerical values of various form factors are tabulated and plotted below for different values of $Q^2$.
\begin{table}[htb]
\begin{tabular}{|c|c|c|c|c|c|c|c|c|} \hline
$Q^2$ $(GeV^2)$ & $C^A_3$ & $C^A_4$ & $C^A_5$ & $C^A_6$ & $G_{\pi N
\Delta}$
\\ \hline
0.15 & 2.492 & 0.0018 & -4.866 & 2.266 & -98
\\ \hline
0.34 & 3.635 & 0.0026 & -8.367 & 3.455 & -169
\\ \hline
0.53 & 3.942 & 0.0027 & -9.749 & 3.668 & -197
\\ \hline
0.71 & 3.880 & 0.0026 & -10.104 & 3.514 & -204
\\ \hline
0.87 & 3.698 & 0.0024 & -10.008 & 3.265 & -202
\\ \hline
1.04 & 3.450 & 0.0022 & -9.677 & 2.964 & -196
\\ \hline
1.34 & 2.983 & 0.0018 & -8.830 & 2,446 & -178
\\ \hline
1.49 & 2.759 & 0.0016 & -8.360 & 2.213 & -169
\\ \hline
1.63 & 2.562 & 0.0015 & -7.920 & 2.014 & -160
\\ \hline
1.77 & 2.377 & 0.0014 & -7.484 & 1.835 & -151
\\ \hline
1.90 & 2.216 & 0.0012 & -7.098 & 1.684 & -143
\\ \hline
2.03 & 2.056 & 0.0011 & -6.714 & 1.549 & -136
\\ \hline
2.15 & 1.939 & 0.0010 & -6.394 & 1.437 & -129
\\ \hline
2.40 & 1.699 & 0.0009 & -5.758 & 1.240 & -116
\\ \hline
\end{tabular}
\caption{}
\end{table}

\newpage
\begin{table}[htb]
\begin{tabular}{|c|c|c|c|c|c|c|c|c|c|c} \hline
$Q^2$ $(GeV^2)$ & $G_1$ & $G_2$ & $G_3$ & $G_{M1}$ & $G_{E2}$ &
$G_{C2}$ & $R_{EM}$ & $R_{SM}$
\\ \hline
0.15 & -12.331 & -0.280 & 390 & -31.681 & -15.537 & 44.997 & -0.490
& 0.224
\\ \hline
0.34 & -15.515 & -0.147 & 205 & -39.060 & -18.232 & 5.983 & -0.467 &
0.036
\\ \hline
0.53 & -14.728 & -0.088 & 122 & -36.060 & -16.630 & -7.040 & -0.500
& -0.056
\\ \hline
0.71 & -13.160 & -0.058 & 81 & -33.232 & -14.545 & -11.288 & -0.438
& -0.116
\\ \hline
0.87 & -11.695 & -0.042 & 58 & -29.564 & -12.566 & -12.378 & -0.425
& -0.159
\\ \hline
1.04 & -10.245 & -0.030 & 42 & -25.951 & -10.694 & -12.288 & -0.412
& -0.196
\\ \hline
1.34 & -8.088 & -0.019 & 26 & -20.875 & -8.281 & -11.100 & -0.397 &
-0.250
\\ \hline
1.49 & -7.198 & -0.015 & 21 & -18.786 & -7.331 & -10.348 & -0.390 &
-0.273
\\ \hline
1.63 & -6.467 & -0.012 & 16.8 & -16.774 & -6.307 & -9.461 & -0.376 &
-0.293
\\ \hline
1.77 & -5.822 & -0.010 & 13.9 & -15.183 & -5.581 & -8.725 & -0.368 &
-0.311
\\ \hline
1.90 & -5.290 & -0.008 & 11.7 & -13.840 & -4.967 & -8.060 & -0.359 &
-0.326
\\ \hline
2.03 & -4.816 & -0.007 & 19.92 & -12.644 & -4.432 & -7.442 & -0.351
& -0.341
\\ \hline
2.15 & -4.424 & 0.006 & 8.58 & -11.662 & -4.003 & -6.917 & -0.343 &
-0.354
\\ \hline
2.40 & -3.724 & 0.005 & 6.42 & -9.892 & -3.243 & -5.934 & -0.328 &
-0.378
\\ \hline
\end{tabular}
\caption{}
\end{table}

\newpage
\hspace{25cm}
\begin{figure}
\includegraphics[width=124mm]{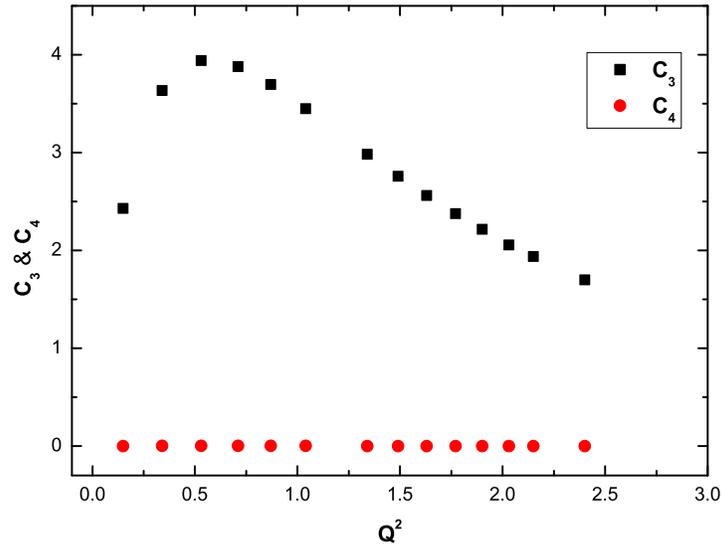}
\caption{This graph shows $C_{3}^{A}(Q^2)$~(black-square) and
$C_{4}^{A}(Q^2)$~(red-circle) as a function of $Q^2$, respectively.
$C_{4}^{A}(Q^2)$ has nearly zero-values for all $Q^2$.}\label{fig1}
\end{figure}

\begin{figure}\centering
\includegraphics[width=124mm]{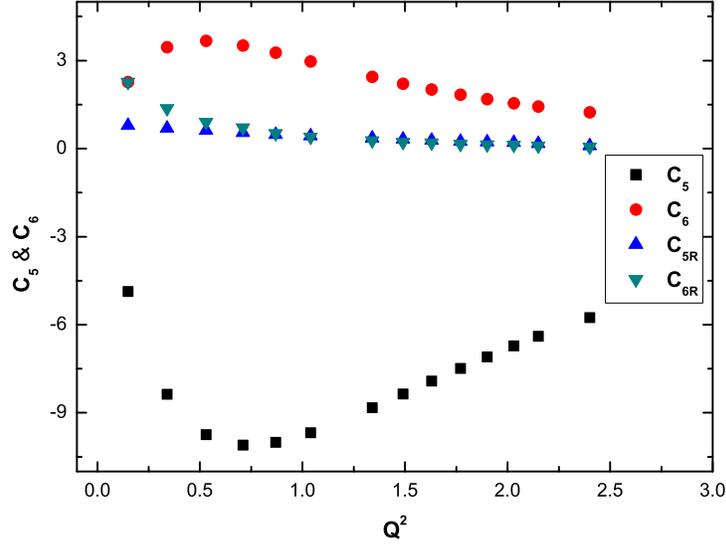}
\caption{The black-square and red-circle represent $C_{5}^{A}(Q^2)$
and $C_{6}^{A}(Q^2)$ as a function of $Q^2$, respectively. The
blue-triangle and green-inverted triangle represent the lattice
results for $C_{5}^{A}(Q^2)$ and $C_{6}^{A}(Q^2)$ of
\cite{Alexandrou:2007zz}.}\label{fig2}
\end{figure}

\begin{figure}\centering
\includegraphics[width=124mm]{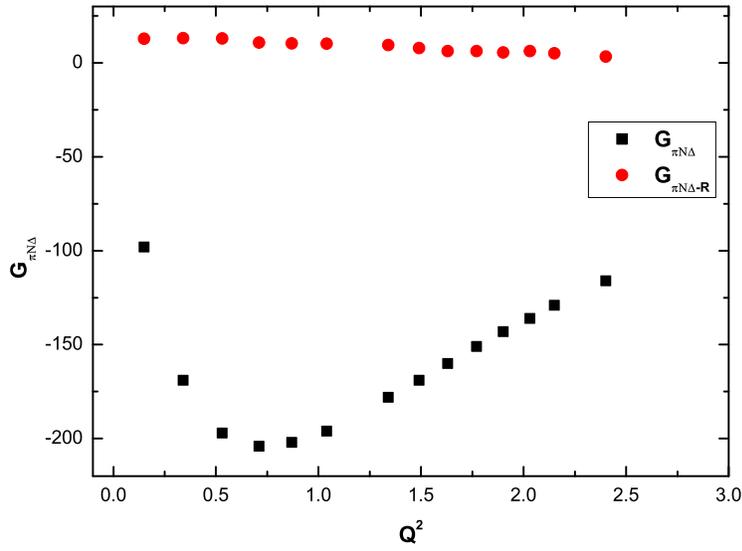}
\caption{The black-square is the numerical results of our model and
the the red-circle is the lattice results of
~\cite{Alexandrou:2007zz} for pion-nucleon-$\Delta$ form factor,
$G_{\pi N\Delta}$.}\label{fig3}
\end{figure}

\begin{figure}
\includegraphics[width=124mm]{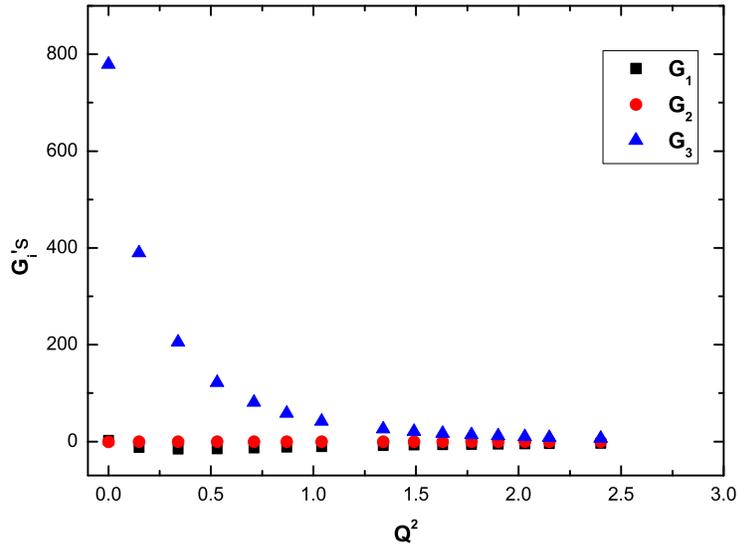}
\caption{This graph shows $G_{i}(Q^2)$ as a function of
$Q^2$.}\label{fig4}
\end{figure}

\begin{figure}\centering
\includegraphics[width=124mm]{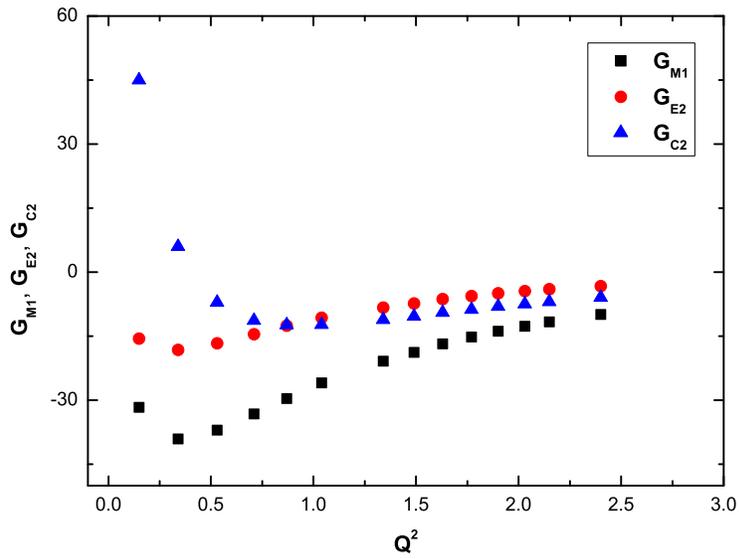}
\caption{The graph of magnetic dipole~(black-square), electric
quadrupole~(red-circle) and Coulomb quadrupole~(blue triangle) form
factors as a function of $Q^2$.}\label{fig5}
\end{figure}

\begin{figure}\centering
\includegraphics[width=125mm]{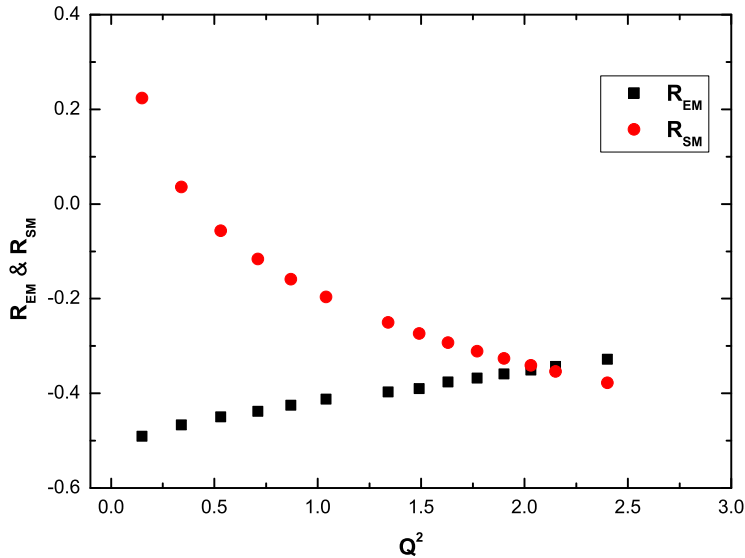}
\caption{The graph of the ratios of the electric~(black-square) and
Coulomb quadrupole~(red-circle) amplitudes to the magnetic dipole
amplitude.}\label{fig6}
\end{figure}

\newpage
\hspace{25cm}

\section{Summary and Conclusions}

The AdS/QCD has provided us a useful tool to calculate hadronic
spectra and meson-baryons transition couplings. We considered the
holographic description of spin 3/2 baryons ($\Delta$ resonaces) in
terms of bulk Rarita-Schwinger fields, which satisfy a set of Dirac equations, upon imposing constraints to project
out their spin-1/2 components.
Then the boundary conditons at the
infra-red cut-off give the free spectra. We considered the effects of
Yukawa couplings and they turn out to be very small for a sizable
value of the coupling constant.

 We also considered meson-baryon transition couplings in the bulk. In particular we considered pion-nucleon couplings. The same term also contribute to four dimensinal rho-nucleon couplings. The numerical results for the meson-baryon couplings from AdS/QCD are in good agreement with the the results obtained from other methods.
The form factors for nucleon and $\Delta$ transitions are also evaluated numerically from AdS/QCD.


\acknowledgments

One of us (D.K.H.) is thankful to M. Procura for a useful discussion.
This work is supported in part by the Korea Research Foundation Grant funded by
the Korean Government (MOEHRD, Basic Research Promotion Fund) (KRF-2007-314-
C00052)(D. K. H.), by KOSEF Basic Research Program with the grant No. R01-2006-000-
10912-0 (D. K. H. and C. P.).


\begin{thebibliography}{99}

\bibitem{Maldacena:1997re}
  J.~M.~Maldacena,
  Adv.\ Theor.\ Math.\ Phys.\  {\bf 2}, 231 (1998)
  [Int.\ J.\ Theor.\ Phys.\  {\bf 38}, 1113 (1999)]
  [arXiv:hep-th/9711200].

\bibitem{Witten:1998qj}
  E.~Witten,
  Adv.\ Theor.\ Math.\ Phys.\  {\bf 2}, 253 (1998)
  [arXiv:hep-th/9802150].
  S.~S.~Gubser, I.~R.~Klebanov and A.~M.~Polyakov,
  Phys.\ Lett.\  B {\bf 428}, 105 (1998)
  [arXiv:hep-th/9802109].

\bibitem{Erlich:2005qh}
  J.~Erlich, E.~Katz, D.~T.~Son and M.~A.~Stephanov,
  Phys.\ Rev.\ Lett.\  {\bf 95}, 261602 (2005)
  [arXiv:hep-ph/0501128].

\bibitem{deTeramond:2005su}
  G.~F.~de Teramond and S.~J.~Brodsky,
  Phys.\ Rev.\ Lett.\  {\bf 94}, 201601 (2005)
  [arXiv:hep-th/0501022].



\bibitem{Son:2003et}
  D.~T.~Son and M.~A.~Stephanov,
  Phys.\ Rev.\  D {\bf 69}, 065020 (2004)
  [arXiv:hep-ph/0304182].
  L.~Da Rold and A.~Pomarol,
  Nucl.\ Phys.\  B {\bf 721}, 79 (2005)
  [arXiv:hep-ph/0501218].

\bibitem{Erlich:2006hq}
  J.~Erlich, G.~D.~Kribs and I.~Low,
  Phys.\ Rev.\  D {\bf 73}, 096001 (2006)
  [arXiv:hep-th/0602110].




\bibitem{Kruczenski:2003uq}
  M.~Kruczenski, D.~Mateos, R.~C.~Myers and D.~J.~Winters,
  JHEP {\bf 0405}, 041 (2004)
  [arXiv:hep-th/0311270].

\bibitem{Sakai:2004cn}
  T.~Sakai and S.~Sugimoto,
  Prog.\ Theor.\ Phys.\  {\bf 113}, 843 (2005)
  [arXiv:hep-th/0412141].

\bibitem{Nawa:2006gv}
  K.~Nawa, H.~Suganuma and T.~Kojo,
  Phys.\ Rev.\  D {\bf 75}, 086003 (2007)
  [arXiv:hep-th/0612187].

\bibitem{Hong:2007kx}
  D.~K.~Hong, M.~Rho, H.~U.~Yee and P.~Yi,
  arXiv:hep-th/0701276.

\bibitem{Hata:2007mb}
  H.~Hata, T.~Sakai, S.~Sugimoto and S.~Yamato,
  arXiv:hep-th/0701280.






\bibitem{Volovich:1998tj}
  A.~Volovich,
  JHEP {\bf 9809}, 022 (1998)
  [arXiv:hep-th/9809009].
  A.~S.~Koshelev and O.~A.~Rytchkov,
  Phys.\ Lett.\  B {\bf 450}, 368 (1999)
  [arXiv:hep-th/9812238].
  R.~C.~Rashkov,
  Mod.\ Phys.\ Lett.\  A {\bf 14}, 1783 (1999)
  [arXiv:hep-th/9904098].
  P.~Matlock and K.~S.~Viswanathan,
  Phys.\ Rev.\  D {\bf 61}, 026002 (2000)
  [arXiv:hep-th/9906077].

\bibitem{Hong:2006ta}
  D.~K.~Hong, T.~Inami and H.~U.~Yee,
  Phys.\ Lett.\  B {\bf 646}, 165 (2007)
  [arXiv:hep-ph/0609270].


\bibitem{Gubser:1998bc}
  S.~S.~Gubser, I.~R.~Klebanov and A.~M.~Polyakov,
  ``Gauge theory correlators from non-critical string theory,''
  %
  Phys.\ Lett.\ B {\bf 428}, 105 (1998).



\bibitem{Stuart:1996zs}
  L.~M.~Stuart {\it et al.},
  Phys.\ Rev.\  D {\bf 58}, 032003 (1998)
  [arXiv:hep-ph/9612416];
  P.~Grabmayr and A.~J.~Buchmann,
  Phys.\ Rev.\ Lett.\  {\bf 86}, 2237 (2001)
  [arXiv:hep-ph/0104203].

\bibitem{Alexandrou:2007zz}
  C.~Alexandrou et al,
  Phys. Rev.\  D {\bf 76}, 094511 (2007).


\bibitem{Alexandrou:2007dt}
  C.~Alexandrou et al,
  Phys. Rev.\  D {\bf 77},085012 (2008)
  [arXiv:0710.4621 [hep-lat]].


\bibitem{Procura:2008ze}
  M.~Procura,
  Phys.\ Rev.\  D {\bf 78}, 094021 (2008)
  [arXiv:0803.4291 [hep-ph]].

\bibitem{Kim:1985ez}
  H.~J.~Kim, L.~J.~Romans and P.~van Nieuwenhuizen,
  Phys.\ Rev.\  D {\bf 32}, 389 (1985).

\bibitem{Hong:2007tf}
  D.~K.~Hong, H.~C.~Kim, S.~Siwach and H.~U.~Yee,
  JHEP {\bf0711}, 036 (2007)
  [arXiv:0709.0314 [hep-ph]].





\bibitem{Riska:2000gd}
  D.~O.~Riska and G.~E.~Brown,
  Nucl.\ Phys.\  A {\bf 679}, 577 (2001)
  [arXiv:nucl-th/0005049].


\bibitem{Jones:1972ky}
  H.~F.~Jones and M.~D.~Scadron,
  Ann. Phys.\  {\bf 81}, 1 (1973).









\end{thebibliography}
\end{document}